\documentclass[aps,prl,superscriptaddress,twocolumn,oneside,floatfix,nofootinbib]{revtex4}

\usepackage{amsmath}
\usepackage{amsfonts}
\usepackage{graphicx}
\usepackage{mathrsfs}
\usepackage{rotating}
\usepackage{setspace}
\usepackage{supertabular}
\usepackage{amsthm}
\usepackage{subfigure}
\usepackage{multirow}

\begin{document}
\title{Phase diagram and structural diversity of the densest binary sphere
packings}

\author{Adam B. Hopkins}
\affiliation{Department of Chemistry, Princeton University, Princeton, New
Jersey 08544, USA}
\author{Yang Jiao}
\affiliation{Princeton Institute for the Science and Technology of Materials,
Princeton University, Princeton, New Jersey 08544, USA}
\author{Frank H. Stillinger}
\affiliation{Department of Chemistry, Princeton University, Princeton, New
Jersey 08544, USA}
\author{Salvatore Torquato}
\affiliation{Department of Chemistry, Princeton University, Princeton, New
Jersey 08544, USA}
\affiliation{Princeton Institute for the Science and Technology of Materials,
Princeton University, Princeton, New Jersey 08544, USA}
\affiliation{Department of Physics, Princeton University, Princeton, New Jersey
08544, USA}
\affiliation{Princeton Center for Theoretical Science, Princeton University,
Princeton, New Jersey, 08544, USA}
\affiliation{Program in Applied and Computational Mathematics, Princeton
University, Princeton, New Jersey 08544, USA}

\begin{abstract}
The densest binary sphere packings have historically been very difficult to
determine. The only rigorously known packings in the $\alpha$-$x$ plane of
sphere radius ratio $\alpha$ and relative concentration $x$ are at the Kepler
limit $\alpha =1$, where packings are monodisperse. Utilizing an implementation
of the Torquato-Jiao sphere-packing algorithm [S. Torquato and Y. Jiao, Phys.
Rev. E {\bf 82}, 061302 (2010)], we present the most comprehensive determination
to date of the phase diagram in $(\alpha,x)$ for the densest binary sphere
packings. Unexpectedly, we find many distinct new densest packings.

\end{abstract}

\maketitle

A packing is defined as a set of nonoverlapping objects arranged in a space of
dimension $d$, and its packing fraction $\phi$ is the fraction of space that the
objects cover. Packings of spheres can be used to describe the structures and
some fundamental properties of a diverse range of substances from crystals and
colloids to liquids, amorphous solids and glasses
\cite{TorquatoRHM2002,CLPCMP1995,HST2011a}. In particular, the
densest sphere packings in $d$-dimensional Euclidean space ${\mathbb R}^d$, or
packings with maximal packing fraction $\phi_{max}$, often correspond to ground
states of systems of particles with pairwise interactions dominated by steep
isotropic repulsion \cite{Pollack1964a,Sanders1980a,BOP1992a,AG2010a}. Recently,
packings of different sized spheres in ${\mathbb R}^3$ have been employed to
model the structures of a range of materials, including, for example, solid
propellants and concrete \cite{KRW2010a,MSJ2008a}. The focus of the present
paper is binary sphere packings, packings of spheres of two sizes, which have
long been used as models for the structures of a wide range of alloys
\cite{SMA1969,Sanders1980a,SVC1982a,EMF1993a}.

Past efforts to identify the densest binary sphere packings have 
employed simple crystallographic techniques \cite{HH2008a,MS1980a} and algorithmic 
methods, {\it e.g.}, Monte Carlo and genetic algorithm \cite{KHH2008a,FD2009a}. 
However, these methods have achieved only limited success, in part due to the very 
large parameter space in $(\alpha,x)$ of binary packings, where 
$\alpha \equiv R_S/R_L$ and $x \equiv \frac{N_S}{N_S+N_L}$, with $R_S$, $N_S$ and 
$R_L$, $N_L$ the respective radii and numbers of the small and large spheres in the 
packing, and where in the infinite volume limit $N_S +N_L \rightarrow \infty$, $x$ 
remains constant. Employing traditional algorithms, difficulties result from the 
enormous number of steps required to escape from local minima in ``energy'', defined 
as the negative of the packing fraction.

In this work, we present the most comprehensive determination to date of the
phase diagram for the densest infinite binary sphere packings. Employing an
algorithmic search using an implementation of the Torquato-Jiao (TJ) linear
programming algorithm \cite{TJ2010a}, we identify $17$ distinct alloys, including 
seven that were heretofore unknown, present in the densest packings over a range of 
$(\alpha,x)$ where significantly fewer were thought to be found. Previously, the 
alloys thought to be present for $\alpha > \sqrt{2} -1 = 0.414213\dots$ 
corresponded to structures such as the AlB$_2$(hexagonal $\omega$), HgBr$_2$, and 
AuTe$_2$ structures \cite{FD2009a,LH1993a,MS1980a}, and to two structures 
composed of equal numbers of small and large spheres \cite{MH2010a}. For 
$\alpha \leq \sqrt{2}-1$, the alloys thought to be present were XY$_n$ structures 
of close-packed large spheres with small spheres (in a ratio of $n$ to 
$1$) in the interstices, {\it e.g.}, the NaCl packing for $n=1$. Using the TJ 
algorithm, we always identify either the densest previously known alloy, or one 
that is denser.

The finding that such a broad array of different densest stable structures
consisting of only two types of components can form without any consideration of
attractive or anisotropic interactions is of significant practical importance.
Our findings strongly suggest that the wide variety of atomic, molecular, and
granular structures found in nature may owe much of their structural diversity
to entropic (free-volume maximizing) interactions rather than only to
anisotropies in nearest-neighbor bonding.

Structures, or configurations of points, can be classified as either periodic or
aperiodic. Roughly defined, a {\it periodic} structure (packing) is one
consisting of a certain number of points (sphere centers), called the {\it
basis}, placed in a defined region of space, the {\it fundamental cell},
replicated many times such that the cells cover all space without any overlap
between cells (or spheres). If a fundamental cell has a {\it minimal basis},
then a smaller cell and basis with the same periodic structure does not exist.
An {\it aperiodic} structure has an infinite minimal basis. We use the term
``alloy'' in a general sense to mean a structure composed of two or more
distinguishable components that are not phase-separated.

The problem of generating dense packings of nonoverlapping nonspherical
particles within an adaptive fundamental cell subject to periodic boundary
conditions has been posed as an optimization problem called the
adaptive-shrinking cell (ASC) scheme \cite{TJ2009a}. The TJ sphere-packing
algorithm \cite{TJ2010a} is a linear-programming solution of the ASC scheme for
the special case of packings of spheres with a size distribution for which
linearization of the design variables, including the periodic simulation box
shape and size, and impenetrability constraints, is exact. The TJ algorithm
leads to strictly jammed packings of spheres with variable degrees of order,
including the maximally dense packings. In a strictly jammed packing, no
volume-decreasing deformation of the fundamental cell or any internal collective
particle motions are possible. Consequently, maximally dense periodic packings
must also be strictly jammed because otherwise the volume of the packing could
be reduced \cite{TS2010a}.

\begin{figure}[ht]
\centering
\includegraphics[angle=270,width = 3.375in,viewport = 160 36 485 727,clip]{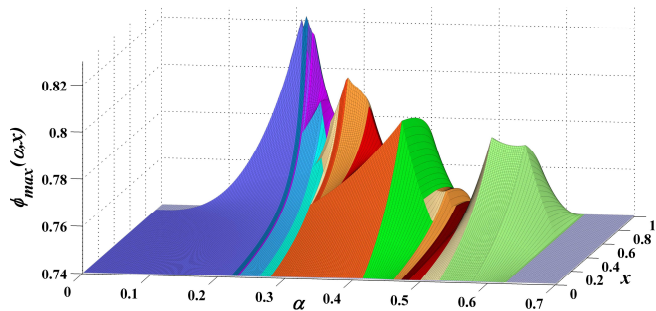}
\caption{(Color online) The most comprehensive determination to date of the
phase diagram and maximal packing fraction surface $\phi_{max}(\alpha,x)$ of the
densest infinite binary sphere packings. The highest point is
$\phi_{max}(0.224744\dots,10/11) = 0.824539\dots$, and all packings for $\alpha
> 0.623387\dots$ consist of two phase-separated monodisperse Barlow phases. We 
have excluded the rectangular region $\alpha < 0.20$, $x > 11/12$. Shadings 
indicate phase composition, as specified in Fig. \ref{phaseDiagram}.}
\label{surface1}
\end{figure}

\begin{figure*}[ht]
\centering
\includegraphics[width = 7.00in,viewport = 12 185 711 397,clip]{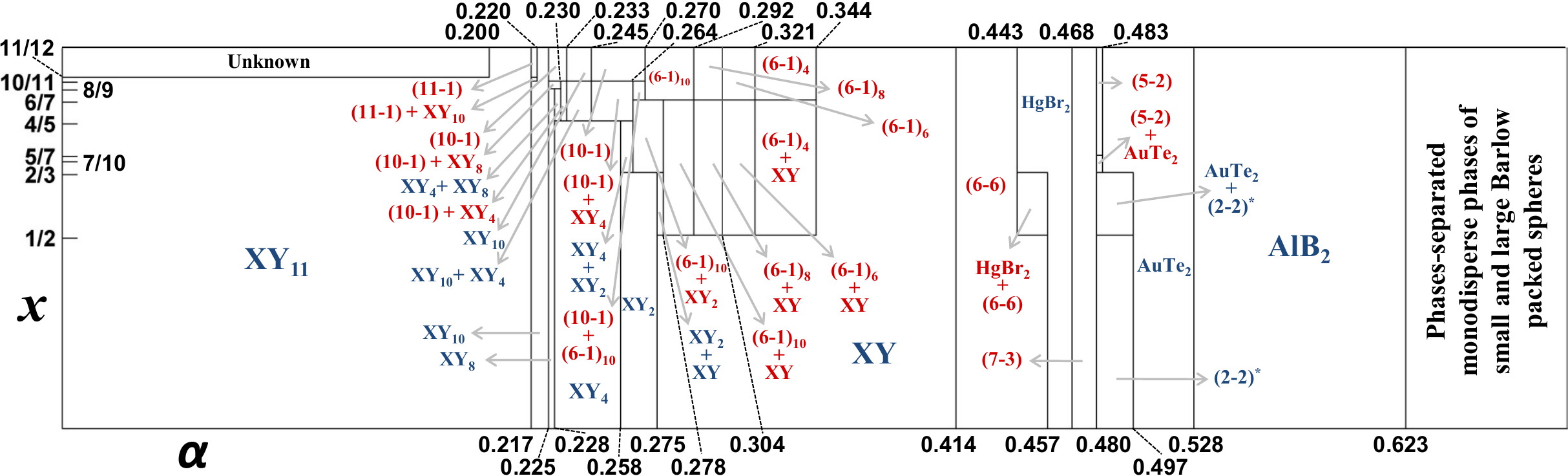}
\caption{(Color online) Phase diagram in $(\alpha,x)$, excluding the region
$\alpha < 0.2$ and $x > 11/12$, of the densest-known infinite binary sphere
packings considering periodic packings with minimal bases of $12$ or fewer
spheres.}
\label{phaseDiagram}
\end{figure*}

Using the TJ algorithm, we have systematically surveyed the parameter space
$(\alpha,x) \in [0,1] \times [0,1]$, omitting the rectangular area $\alpha <
0.2$, $x > 11/12$ for reasons mentioned below, to find the putative densest
binary packings for all bases of up to $12$ spheres. From this survey, we construct
the most comprehensive determination to date of the phase diagram of the densest
infinite binary packings, and the best-known lower bound on the function 
$\phi_{max}(\alpha,x)$, the packing fraction of the densest infinite packings 
of binary spheres at fixed $(\alpha,x)$ for the values of $\alpha$ in our survey; 
see Fig. \ref{surface1}. We present a detailed view of the composition of phases 
in Figure \ref{phaseDiagram}.

Away from the point $(\alpha,x) = (0,1)$, assuming that the packing fraction 
and composition of the generally small number of densest alloys at specified 
radius ratio $\alpha$ are known, the densest infinite packings are 
phase-separated combinations of alloy and/or monodisperse phases. The spheres 
in the monodisperse phases are packed as any of the uncountably infinite 
number of Barlow packings \cite{Barlow1883a}, {\it e.g.}, the well known fcc 
and hcp close-packed packings. However, as $\alpha\rightarrow 0$ and 
$x\rightarrow 1$, the number of distinct densest packings approaches infinity 
due to the infinite number of XY$_n$ and similar packings. For this reason, 
we exclude the region $\alpha < 0.2$, $x > 11/12$ from our study, truncating 
at $\alpha = 0.2$ because it is close to the maximum value $\alpha=0.216633\dots$ 
at which $11$ small spheres fit in the interstices of a Barlow packing of large 
spheres.

When $\alpha$ is near the Kepler limit of unity, the densest packings consist of
two phase-separated monodisperse Barlow phases of small and large spheres with
packing fraction $\pi/\sqrt{18}$ \cite{TS2010a}. This is the case in Fig.
\ref{surface1} for all packings with $\alpha > 0.623387\dots$. In general, the 
surface is continuous and piecewise differentiable, though as 
$\alpha\rightarrow 0$ and $x\rightarrow 1$, the density of curves along which 
the surface is not differentiable approaches infinity.

In ${\mathbb R}^2$, periodic, quasicrystalline\footnotemark[1], directionally
periodic\footnotemark[2] and disordered\footnotemark[3] structures can all be
found among the putative densest binary disk packings
\cite{LH1993a,TothRF1964,LHC1989a}. We believe that all of these types of
structures might be present among the densest binary sphere packings in
${\mathbb R}^3$ as well, though we do not identify quasicrystalline or disordered
structures here. Due to computational constraints attributable to the
scope and resolution of our survey in $(\alpha,x)$, we have limited our
investigation to periodic packings considering all possible bases of $12$ and 
fewer spheres. This limitation substantially increases the difficulty of identifying 
any aperiodic packings, which most often cannot be approximated well by a periodic 
packing with a basis of $12$. The directionally periodic packings 
that we have identified are those for which no boundary cost exists between phases, 
{\it e.g.}, between AlB$_2$ and monodisperse phases, and these packings are
therefore degenerate in density with periodic packings.

\footnotetext[1]{A quasicrystal can be roughly described as an aperiodic
structure that nonetheless exhibits bond orientational order in symmetries ({\it
e.g.}, icosahedral) forbidden to periodic crystals.}
\footnotetext[2]{We describe a directionally periodic structure as an aperiodic
structure that exhibits a period along at least one spatial axis but never
simultaneously along all $d$ vectors that span the space.}
\footnotetext[3]{For the purposes of this paper, a disordered structure is
aperiodic but neither a quasicrystal nor directionally periodic.}


{\it Identifying the densest packings:}
To identify the densest infinite binary packings in ${\mathbb R}^3$, we begin
with the obvious statement that at all given $(\alpha,x)$, there is a densest
packing that consists of a finite number of phase-separated alloy and
monodisperse phases. Since we limit ourselves in this work to minimal bases of
no more than $12$ spheres, we must assume that all of these alloy phases can be
constructed from repetitions of local structures consisting of $12$ spheres or
fewer. Though we recognize that this latter assumption is most likely false for
some values of $(\alpha,x)$, we contend that for the majority of the area of the
parameter space studied, it is correct.

We describe a distinct alloy as one with a unique combination of composition of
spheres in its minimal basis and lattice system characterization of its
fundamental cell. This is a more encompassing characterization than that applied
in Ref. \cite{LH1993a}, where periodic alloys in the densest binary disk
packings were classified by composition and the numbers of small and large
sphere contacts in the fundamental cell. For example, the distinct alloy with 
six small and one large sphere in its minimal basis and fundamental cell belonging 
to the triclinic lattice system exhibits a wide range of contact networks over the 
range $0.292 \leq \alpha \leq 0.344$ when it appears in the densest packings. To 
illustrate this point, we have divided this alloy into three subcategories, 
($6$-$1$)$_8$, ($6$-$1$)$_6$, and ($6$-$1$)$_4$, where the subscript indicates the 
number of large sphere contacts per large sphere, as depicted in the detailed phase 
diagram (Fig \ref{phaseDiagram}). We note that the alloy could be further 
subdivided by the numbers of small-small and large-small contacts.

The boundaries between phases are negligible in the infinite volume limit, and so 
the packing fraction of a collection of phase-separated monodisperse and $\beta$ 
distinct alloy phases can be written as,
\begin{equation}
\phi(\alpha,x) = \frac{\frac{4\pi}{3}\left((1\!-\!x)R_L^3 \!+\! xR_S^3\right)}{xC_{B}^S \!+\!
(1\!-\!x)C_{B}^L \!+\! \sum_{i=1}^{\beta}\!x_i^L\!\left(\frac{C_i}{L_i} \!-\!
\frac{S_i}{L_i}C_{B}^S \!-\! C_{B}^L\right)},
\label{packingFraction1}
\end{equation}
with $C_{B}^S$ and $C_{B}^L$ the volume per sphere, respectively, in a
close-packed Barlow packing of small and large spheres, $x_i^L$ the relative 
fraction of large spheres distributed in alloy phase $i$, and $C_i$ the volume of 
a fundamental cell of alloy phase $i$ containing $L_i$ large and $S_i$ small 
spheres. The constraints $x_i^L \geq 0$, $\sum_{i=1}^{\beta}x_i^L \leq 1-x$, and 
$\sum_{i=1}^{\beta}(S_i/L_i)x_i^L \leq x$ must be valid due to conservation of 
total particle numbers.

To find the densest packing $\phi_{max}(\alpha,x)$ from among $\beta$ alloy and
two close-packed monodisperse phases, Eq. (\ref{packingFraction1}) must be
maximized. This is accomplished by treating the aforementioned $\beta + 2$
constraints and the summation term in the denominator of Eq.
(\ref{packingFraction1}) as a linear programming problem where the objective is
to minimize the summation. From this postulation, it can be
proved \cite{HST2011b} that there is always a densest binary packing consisting
of no more than two phase-separated phases, though it may be degenerate in
density with packings consisting of more than two phases or with mixed phase
packings.

Using Eq. (\ref{packingFraction1}) and considering a fixed value of $\alpha$,
the densest infinite binary packings constructed from binary alloys with bases
of $12$ or fewer spheres can be found for all values of $x$. This only requires
knowing the densest packings in a fundamental cell for combinations of positive
integer $S_i$ and $L_i$ such that $S_i + L_i = 2,3,\dots,12$. Employing the TJ
algorithm, we have solved these problems (putatively) to accuracy of about
$10^{-4}$ in $\phi$ for $\alpha$ spaced $0.025$ apart, and on a finer grid with
$\alpha$ spaced about $0.0028$ apart for certain values of $S_i$ and $L_i$ where
particularly dense packings were identified.

Figure \ref{phaseDiagram} is our determination of the phase diagram, described
with heretofore unattained accuracy, for the densest infinite binary sphere
packings considering periodic packings with minimal bases of $12$ or fewer
spheres. In the diagram, the seven previously unrecognized distinct alloys are
described according to the composition of their minimal basis, {\it e.g.},
($6$-$6$) for a packing with $6$ small and $6$ large spheres per fundamental
cell. In Fig. \ref{phaseDiagram}, the points (lines) where the composition 
of phase-separated phases changes from alloy plus monodisperse packing of small
spheres to the same alloy plus large spheres are not drawn. Additionally, when
only one alloy is listed, it is assumed that the densest packing consists of a
monodisperse phase and an alloy phase, except at points such that $x =
S_i/(S_i+L_i)$, with $S_i$ and $L_i$ the respective numbers of small and large
spheres in the alloy phase listed, where only the alloy phase is present.

We briefly describe the $17$ distinct alloys here, leaving the detailed 
descriptions for a later work \cite{HST2011b}. The XY$_n$ alloys are present for 
$n=1$, $2$, $4$, $8$, $10$ and $11$. In these packings, the large spheres are 
close-packed Barlow packings, and the small spheres are inside the interstices 
as rattlers, movable but caged spheres, except for at ``magic'' \cite{LH1993a} 
$\alpha$ where they are jammed. Additionally, for $n=2$, $4$, $8$ and $10$, 
there are XY$_n$ alloys for $\alpha$ greater than the magic radius ratios. These 
packings consist of large spheres arranged as in a Barlow packing but not in 
contact, with interstitial jammed small spheres arranged as was the case for the 
magic $\alpha$.

The AlB$_2$ alloy is well known, and the HgBr$_2$ and AuTe$_2$ 
alloys, described in another work \cite{FD2009a}, have, respectively, four small 
and two large and two small and one large spheres in their fundamental cells. The 
alloys belong to the orthorhombic and monoclinic lattice systems, respectively. The 
alloy listed as ($2$-$2$)$^*$ exhibits the same packing fraction (error of 
less than $10^{-4}$) over the range $0.480 \leq \alpha \leq 0.497$ where it appears 
in the densest packings as the alloy described in Ref. \cite{MH2010a} as 
``Structure 2'', though ``Structure 2'' has four small and four large spheres 
in its fundamental cell. Due to the precise agreement of packing fractions, we 
postulate that the two alloys may be the same or have materially negligible 
differences over the range $0.480 \leq \alpha \leq 0.497$.

There are two alloys, ($11$-$1$) and ($10$-$1$), that are similar to the
XY$_{11}$ and XY$_{10}$ packings except that their fundamental cells belong to
the tetragonal and rhombohedral lattice systems, respectively, as opposed to
cubic. The ($6$-$1$)$_{10}$ alloy can be described as an orthorhombic body-centered 
packing of large spheres, each with ten large-large sphere contacts, with four 
small spheres on each face. The alloy subdivided as ($6$-$1$)$_8$, ($6$-$1$)$_6$, 
and ($6$-$1$)$_4$ in Fig. \ref{phaseDiagram} is similar but with a skewed 
fundamental cell belonging to the triclinic lattice system.

Over the range $0.414 < \alpha < 0.457$ where the ($6$-$6$) alloy appears in the 
densest packings, we have found that in simulation, increasing the basis from one 
small and one large spheres up to six small and six large spheres in a one-to-one
ratio results in alloys with increasing packing fraction. The simulations with 
four large and four small spheres in the fundamental cell produce an alloy with 
packing fractions that agree (error of less than $10^{-4}$) with those of 
``Structure 1'' described in Ref. \cite{MH2010a}. The ($5$-$2$) alloy is arranged 
as offset square lattice layers of large spheres with small spheres in between; 
the alloy belongs to the monoclinic lattice system. The ($7$-$3$) alloy fundamental 
cell belongs to the orthorhombic lattice system. The alloy is similar to three 
adjacent fundamental cells of an AlB$_2$ packing with one extra small sphere 
inserted.

Our determination of the phase diagram (Figs. \ref{surface1} and
\ref{phaseDiagram}) describes an unexpected diversity in the densest binary
sphere packings, with $17$ distinct alloys present. One implication of these 
findings is that entropic (free-volume maximizing) particle interactions 
contribute to the structural diversity of mechanically stable and ground-state 
structures of atomic, molecular, and granular solids. Additionally,
the structures we have identified can be useful as known points of departure
when investigating experimentally the properties of binary solids composed of
particles that exhibit steep isotropic pair repulsion. Finally, our results serve 
as crucial references states for studies of corresponding disordered binary 
sphere packings.

We have carried out a comprehensive study of the densest infinite binary sphere
packings at high resolution in $\alpha$ and $x$, leading to the discovery
employing the TJ algorithm of many heretofore unknown structures. Though we have
limited ourselves to minimal bases of $12$ or fewer spheres, the discovery of the 
($7$-$3$), ($6$-$6$), and ($5$-$2$) alloys suggests that periodic structures with 
minimal bases larger than $12$, further directionally-periodic, quasicrystalline 
and disordered structures might be present among the densest packings. Additionally, 
it is possible that densest packings have a relation to the structures of glassy 
binary sphere solids and/or to those of binary sphere liquids near the freezing 
point. In future work, we will investigate these possibilities.

This work was supported by the NSF under award numbers DMR-0820341 and DMS-0804431.



\end{document}